\documentclass{elsart}
\usepackage{graphicx}

\begin{document}

\begin{frontmatter}

\title{Vacancy-driven magnetocaloric effect in Prussian blue analogues}

\author[s3]{Marco Evangelisti\corauthref{cor1}},
\ead{evange@unimore.it}\corauth[cor1]{Corresponding author. Tel.: + 39-059-2055313.}
\author[s3]{Esp\'{e}ran\c{c}a Manuel},
\author[s3,mod]{Marco Affronte},
\author[par]{Masashi Okubo}$^{,\dagger}$,
\author[par]{Cyrille Train},
\author[par]{Michel Verdaguer}
\address[s3]{National Research Center on ``nanoStructures and bioSystems at Surfaces'' (S$^{3}$), INFM-CNR,
41100 Modena, Italy}
\address[mod]{Dipartimento di Fisica, Universit\`{a} di Modena e Reggio Emilia, 41100 Modena, Italy}
\address[par]{Chimie Inorganique et Mat\'eriaux Mol\'eculaires, Unit\'e CNRS 7071, Universit\'e Pierre et
Marie Curie, 75252 Paris, France}

\begin{abstract}
We experimentally show that the magnetocaloric properties of molecule-based Prussian blue analogues can be
adjusted by controlling during the synthesis the amount of intrinsic vacancies. For
Cs$_{x}$Ni$^{II}_{4}$[Cr$^{III}$(CN)$_{6}]_{(8+x)/3}$, we find indeed that the ferromagnetic phase transition
induces significantly large magnetic entropy changes, whose maxima shift from $\sim 68$~K to $\sim 95$~K by
varying the number of [Cr$^{III}$(CN)$_{6}]^{3-}$ vacancies, offering an unique tunability of the
magnetocaloric effect in this complex.
\end{abstract}

\end{frontmatter}

Magnetic ordering phenomena are efficiently exploited to enhance the magnetocaloric effect (MCE) of magnetic
materials~\cite{gschneidner05}. This is possible because the response to the application or removal of
magnetic fields is indeed maximized near the ordering temperature. In the search of suitable materials for
magneto-cooling applications, however, one may need to adjust the ordering temperature to make optimum use of
the magnetocaloric properties of a given material. For conventional materials, such as lanthanide compounds
and alloys, it is common practice, in this respect, to partly substitute one constituting element for another
one~\cite{gschneidner05}. Here we show that a similar but different strategy can be employed as well in
Prussian blue analogues (PBA), which were recently investigated for their magnetocaloric
properties~\cite{manuel06}. The field of magnetocaloric research on molecule-based materials, such as PBA, is
relatively young, although it has already shown promising potentialities~\cite{evange06}.

\begin{figure}[t!]
\centering\includegraphics[angle=0,width=6.9cm]{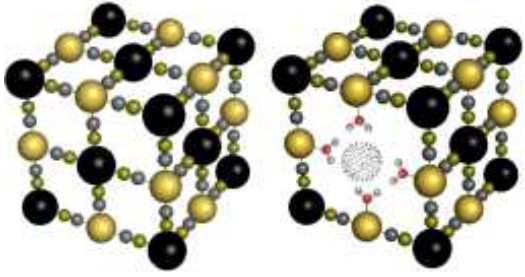} \caption{Sketch of Ni-Cr PBA without any vacancy
(left) and with a [Cr$^{III}$(CN)$_{6}]^{3-}$ vacancy, shown as a large bright sphere, coordinated by bound
water molecules (right). Black spheres represent Cr$^{III}$ ($S_{\rm Cr}=3/2$), whereas lighter-colored
spheres represent Ni$^{II}$ (spin $S_{\rm Ni}=1$), and small circles denote the cyano-bridge that ensures the
exchange coupling between the metallic centers.} \label{fig1}
\end{figure}

The Prussian blue analogues here reported have the idealized formula
Cs$_{x}$Ni$^{II}_{4}$[Cr$^{III}$(CN)$_{6}]_{(8+x)/3}$, and the conventional unit cell is depicted in
Fig.~$\ref{fig1}$. Depending on the value of $x$, the presence of the intrinsic [Cr$^{III}$(CN)$_{6}]^{3-}$
vacancies and their amount per cell is easily seen, as exemplified in the Figure. The non-stoichiometry is
known to be essential for the observation of peculiar phenomena in this class of materials, such as
photomagnetism for Co-Fe PBA~\cite{kawamoto01}. In what follows, we shall focus on two Ni-Cr PBA having $x=0$
and $x=4$, which we shortly denote hereafter as NiCr$_{2/3}$ and CsNiCr, respectively. The difference between
the two resides in the ideally perfect stoichiometry of CsNiCr, for which half of the tetrahedral
interstitial sites are occupied by Cs cations which maintain charge neutrality, and in the absence of Cs in
NiCr$_{2/3}$ that results in the presence of vacancies. Both compounds are known to undergo a transition to a
long-range ferromagnetic ordered state~\cite{gadet92}. Further information on the structure together with a
description of the method of synthesis can be found in Ref.~\cite{gadet92}. Susceptibility and magnetization
measurements were carried out in a commercial apparatus for the $0<H<7$~T magnetic field range. All data were
collected on powdered samples of the compounds.

\begin{figure}[t!]
\centering\includegraphics[angle=0,width=6.9cm]{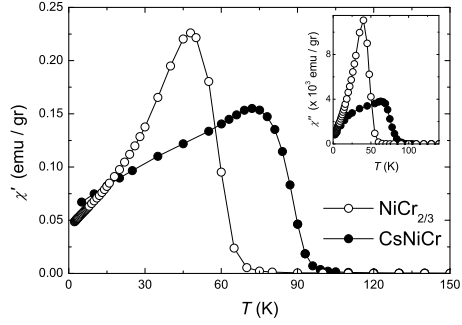} \caption{Complex susceptibility collected at
$f=1730$~Hz and ac-field $h_{ac}=10$~G, for CsNiCr and NiCr$_{2/3}$, as labelled. Inset: out-of-phase
component $\chi^{\prime\prime}(T)$.} \label{fig2}
\end{figure}

For both compounds, Figure~$\ref{fig2}$ shows the complex susceptibility collected with an ac-field
$h_{ac}=10$~G at $f=1730$~Hz. For CsNiCr, the abrupt change of the in-phase susceptibility $\chi^{\prime}(T)$
at $T_{C}\simeq 90$~K is ascribed to the transition to a ferromagnetically ordered state, which is also
corroborated by recent specific heat experiments~\cite{manuel06}. Figure~$\ref{fig2}$ shows as well that
fluctuations in the ordering process gives rise to an out-of-phase susceptibility $\chi^{\prime\prime}(T)$
signal. The complex susceptibility of NiCr$_{2/3}$ follows qualitatively that of CsNiCr but at much lower
temperatures with a break in $\chi^{\prime}(T)$ at about 60~K, which is accompanied by a
$\chi^{\prime\prime}(T)$ signal (Fig.~$\ref{fig2}$). The difference between the susceptibilities of the two
compounds can be easily explained within the frame of the molecular field theory for which the $T_{C}$ values
of the present system can be expressed as

\begin{eqnarray}
\nonumber T_{C} & = & \frac{2(z_{\rm Ni}z_{\rm Cr})^{1/2}|J|}{3k_{B}}\\
&& \times [S_{\rm Ni}(S_{\rm Ni}+1)S_{\rm Cr}(S_{\rm Cr}+1)]^{1/2} \label{eq1}
\end{eqnarray}

where $z_{\rm Ni}$ and $z_{\rm Cr}$ are the numbers of nearest neighbour metal ions of the Ni$^{II}$ and
Cr$^{III}$ ions, respectively, and $J$ is the exchange coupling constant between the Ni$^{II}$ and Cr$^{III}$
ions. The inclusion of [Cr$^{III}$(CN)$_{6}]^{3-}$ vacancies accounts for the different $T_{C}$'s, since it
causes a change in the number of nearest neighbours. It is easy to show, indeed, that although $z_{\rm Ni}$
retains its value by switching from CsNiCr to NiCr$_{2/3}$ (for which $z_{\rm Ni}=6$), the $z_{\rm Cr}$
values are 6 and 4 for CsNiCr and NiCr$_{2/3}$, respectively. Given $S_{\rm Ni}=1$, $S_{\rm Cr}=2$ and
assuming $T_{C}=90$~K as for CsNiCr, we obtain from Eq.~($\ref{eq1}$) the exchange coupling whose estimate
amounts to $J\simeq 6.5$~K, where the positive sign is set by the ferromagnetic nature of the ordered phase.
By switching from CsNiCr to NiCr$_{2/3}$, it follows, according to Eq.~($\ref{eq1}$), a lower value for
$T_{C}$ that amounts to $\sim 60$~K for the latter, in good agreement with the susceptibility data
(Fig.~$\ref{fig2}$). In Prussian blue analogues, the dependence of $T_{C}$ on the number of neighbours is
well-known, as was already reported for several derivatives including
Cs$_{x}$Ni$^{II}_{4}$[Cr$^{III}$(CN)$_{6}]_{(8+x)/3}$ as
well~\cite{gadet92,griebler82,verdaguer93,ohkoshi97}.

\begin{figure}[t!]
\centering\includegraphics[angle=0,width=6.9cm]{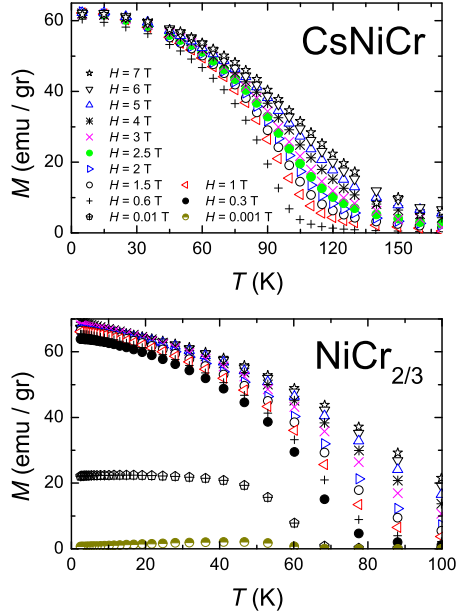} \caption{Field-cooled $M(T)$ curves measured at
different applied-fields for CsNiCr (top) and NiCr$_{2/3}$ (bottom), as labelled in the top panel.}
\label{fig3}
\end{figure}

For a proper evaluation of the MCE of these compounds~\cite{pecharsky99JAP}, we performed systematic
magnetization $M(T,H)$ measurements as a function of temperature and field. Field-cooled $M(T,H)$
measurements for several applied-fields $H$ up to 7~T show spontaneous magnetization below the corresponding
$T_{C}$'s (Fig.~3). In an isothermal process of magnetization, the magnetic entropy change $\Delta S_{m}$ can
be derived from Maxwell relations by integrating over the magnetic field change $\Delta H=H_{f}-H_{i}$, that
is:

\begin{equation}
\Delta S_{m}(T)_{\Delta H}=\int_{H_{i}}^{H_{f}}\frac{\partial M(T,H)}{\partial T}~{\rm d}H.
\end{equation}

\begin{figure}[h!]
\centering\includegraphics[angle=0,width=6.9cm]{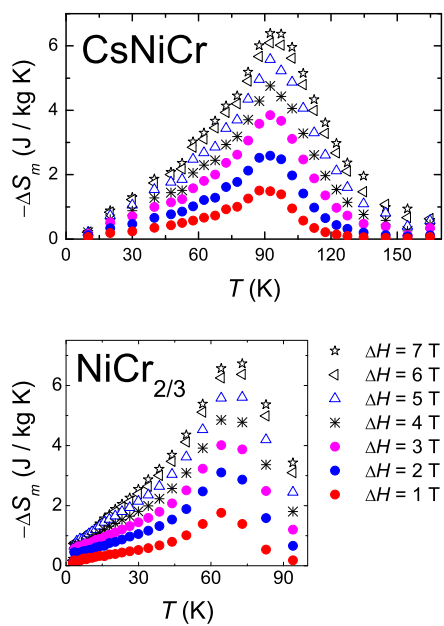} \caption{Magnetic entropy change $\Delta S_{m}(T)$ as
obtained from $M(T,H)$ data of Fig.~3 for CsNiCr (top) and NiCr$_{2/3}$ (bottom), for several field changes
$\Delta H$, as labelled.} \label{fig4}
\end{figure}

From the $M(H)$ data of Fig.~$\ref{fig3}$, the obtained $\Delta S_{m}(T)$ for several $\Delta H$
values~\cite{note3} are depicted in Fig.~$\ref{fig4}$. We note that the maximum change of the magnetic
entropy upon application of a magnetic field, provides values that are similar for both compounds. Indeed, it
can be seen that $-\Delta S_{m}$ increases by increasing $\Delta H$, reaching for $\Delta H=7$~T the values
of 6.6~J~kg$^{-1}$K$^{-1}$ and 6.9~J~kg$^{-1}$K$^{-1}$ for CsNiCr and NiCr$_{2/3}$, respectively. However,
since these changes are associated with the mechanism of magnetic ordering~\cite{manuel06}, it turns out that
they take place at well-separated temperatures: $\sim 95$~K and $\sim 68$~K, for CsNiCr and NiCr$_{2/3}$,
respectively (Fig.~$\ref{fig4}$). Concluding, the possibility of controlling during the synthesis the number
of vacancies, provides an excellent opportunity to shift the MCE of such complexes.

This work is partially funded by the Italian MIUR under FIRB project No. RBNE01YLKN and by the EC-Network of
Excellence ``MAGMANet'' (No. 515767). E.M. acknowledges support from the EC-Marie Curie network ``QuEMolNa''
(No. MRTN-CT-2003-504880). M.O. was supported by the Japan Society for the Promotion of Science for Young
Scientists.


\begin{thebibliography}{00}
\bibitem[$\dagger$]{byline} Present address: National Institute of Advanced Industrial
Science and Technology (AIST), Umezono 1-1-1, Tsukuba, Ibaraki 305-8578, Japan.
\bibitem{gschneidner05} See, e.g., K.A. Gschneidner Jr., V.K. Pecharsky, A.O. Tsokol, Rep. Prog. Phys. 68
(2005) 1479, and references therein.
\bibitem{manuel06} E. Manuel, M. Evangelisti, M. Affronte, M. Okubo, C. Train, M. Verdaguer, Phys. Rev. B
73 (2006) 172406.
\bibitem{evange06} For a recent review, see, e.g., M. Evangelisti, F. Luis, L.J. de Jongh, M. Affronte,
J. Mater. Chem. (to be published), also cond-mat/0603368, and references therein.
\bibitem{kawamoto01} T. Kawamoto, Y. Asai, S. Abe, Phys. Rev. Lett. 86 (2001) 348.
\bibitem{gadet92} V. Gadet, T. Mallah, I. Castro, M. Verdaguer, J. Am. Chem. Soc. 114 (1992) 9213.
\bibitem{griebler82} W.D. Griebler and D. Babel, Z. Naturforsch. B 87 (1982) 832.
\bibitem{verdaguer93} M. Verdaguer, T. Mallah, V. Gadet, I. Castro, C. H\'{e}lary, S. Thi\'{e}baut,
P. Veillet, Conf. Coord. Chem. 14 (1993) 19.
\bibitem{ohkoshi97} S. Ohkoshi, O. Sato, T. Iyoda, A. Fujishima, K. Hashimoto, Inorg. Chem. 36 (1997) 268;
S. Ohkoshi, T. Iyoda, A. Fujishima, K. Hashimoto, Phys. Rev. B 56 (1997) 11642; S.-i. Ohkoshi and K.
Hashimoto, Chem. Phys. Lett. 314 (1999) 210.
\bibitem{pecharsky99JAP} V.K. Pecharsky and K.A. Gschneidner Jr., J. Appl. Phys. 86 (1999) 565.
\bibitem{note3} For practical reasons, the measurements at the lowest applied field were carried
out for $H_{i}=10^{-3}$~T, which in our calculations was approximated to zero-applied-field.
\end{thebibliography}
\end{document}